\documentclass[aps,showpacs,nofootinbib,notitlepage,12pt]{revtex4-1}
\usepackage[pdftex]{graphicx}
\usepackage{dcolumn}
\usepackage{bm}
\usepackage{latexsym}
\usepackage{amsmath}
\usepackage{amsfonts}
\usepackage{amssymb}
\usepackage{color}
\usepackage{array}
\usepackage{framed}
\usepackage{esvect}
\usepackage[latin9]{inputenc}
\usepackage{amsmath}
\usepackage{dsfont}
\usepackage{amstext}
\usepackage{amssymb}
\usepackage{amsbsy}
\usepackage{amsthm}
\usepackage{epsfig}
\usepackage{graphicx}
\usepackage{bbm}
\usepackage{hyperref}

\renewcommand{\>}{\rangle}

\renewcommand{\v}[1]{\mathbf{#1}} 

\newcommand{\T}{\mathcal{T}}
\newcommand{\bpm}{\begin{pmatrix}}
\newcommand{\epm}{\end{pmatrix}}
\newcommand{\be}{\begin{equation}}
\newcommand{\ee}{\end{equation}}
\newcommand{\bea}{\begin{eqnarray}}
\newcommand{\eea}{\end{eqnarray}}

\newcommand{\Z}{\mathbb{Z}}

\begin{document} 




\title{ Symmetry Protected Topological phases of Quantum Matter}
\author{T. Senthil}
\affiliation{Department of Physics, Massachusetts Institute of Technology,
Cambridge, MA 02139, USA}
 \date{\today}
\begin{abstract}
We describe recent progress in our understanding of the interplay between interactions, symmetry,  and topology in states of quantum matter.  We focus on  a minimal generalization of the celebrated topological band insulators to interacting many particle systems, known as Symmetry Protected Topological (SPT) phases.  In common with the topological band insulators these states have a bulk gap and no exotic excitations but have non-trivial surface states that are protected by symmetry.  We describe the various possible such phases and their properties in three dimensional systems with realistic symmetries.  We develop many key ideas of the theory of these states using simple examples. The emphasis is on physical rather than mathematical properties.  We survey insights obtained from the study of SPT phases for a number of other theoretical problems.

\end{abstract} 

\maketitle

\tableofcontents

\section{Introduction}
Following the theoretical prediction and experimental discovery of topological insulators in the last decade\cite{km05,bhz06,mb07,fkm07,rroy,mlcmp,dhsieh,TIs}, attention has turned to 
describing similar topological phenomena in strongly correlated electronic materials.  Experimentally a number of such correlated materials are currently being explored as candidate topological insulators. These include mixed-valent materials\cite{tki} like $SmB_6$ as well as iridium oxide materials on pyrochlore lattices\cite{irreview}. 
On the theoretical side the study of topological insulation in the presence of strong correlations poses fresh challenges.  It requires us to move away from the  crutch of free fermion Hamiltonians and band topology that has thus far informed much of the discussion of the phenomenon of topological insulation. 

In thinking about the interplay of strong correlation and topological phenomena, a number of questions immediately present themselves. Are free fermion topological  phases {\bf stable} to the inclusion of electron interactions? 
Perhaps more interestingly are there new kinds of topological insulators that {\bf require} interactions and have no free fermion counterpart? How many phases are there and what are their physical properties?

How should we generalize the concept of a topological insulator to interacting systems?   
In this review we will describe recent dramatic progress in understanding a {\em minimal} generalization to phases of quantum matter known as Symmetry Protected Topological (SPT) states.  Consider a system of interacting electrons with some definite symmetries.  We require that the bulk of the system have a unique ground state that preserves the symmetries and has a gap to all excitations. Related to this we require that there are no `exotic' excitations which carry, say, fractional quantum numbers  or fractional statistics.  Such a system may nevertheless have non-trivial surface states that are protected by the symmetries. These properties define the concept of an SPT phase of electrons. Clearly the topological band insulators are special cases of SPT phases. 

It is important right away to distinguish SPT states from other more exotic generalizations. For instance in the so-called Fractional Topological Insulators\cite{fTI2d,sbmsfTI3d,jmqietalfTI,nprt,ftimdls,levinstern} the bulk may be gapped but may develop what is known as  intrinsic `topological order' \cite{Wenbook} familiar from studies of the fractional quantum Hall and quantum spin liquid phases. In contrast to topological band insulators and other SPT phases, topological ordered phases have exotic bulk excitations (e.g., with fractional statistics and possibly fractional quantum numbers) and a degenerate ground state on closed manifolds.  An even more exotic generalization is to 
phases where the bulk has gapless excitations (such as the ``Topological Mott Insulator" described in Ref. \cite{pesinlb}).  A very useful perspective on these different generalizations is provided by the structure of many-body quantum entanglement in the ground state of these various phases. In SPT phases - in common with Topological Band Insulators -  the degrees of freedom in one region of a sample are only entangled quantum mechanically with neighboring regions. We may call this Short Range Entanglement (SRE). In contrast, topologically ordered phases and their gapless cousins have what may be called Long Range Entanglement (LRE). 

A classic example of an interacting SPT phase is the Haldane spin chain in one dimension. This has a unique gapped bulk ground state but develops dangling spin-$1/2$ moments in the presence of open boundaries. Modern work on interacting SPT phases have their origins in formal mathematical  classifications\cite{1dsptclass} of all such phases in $d = 1$. In an important advance 
Chen et al\cite{chencoho2011} proposed a generalization of this formal classification to bosonic systems in $d > 1$, based on the concept of group cohomology.  They also provided exactly soluble (albeit rather complex) models for many of these new bosonic SPT phases  in $d > 1$.

This work left a number of fundamental questions open. First the formal methods employed for the classification do not directly shed light on the physics of these phases. 
Second these methods are hard to generalize to electronic systems (in $d > 1$).  A generalization - known as the group super-cohomology\cite{scoho} -  has been attempted but 
cannot handle the Kramers structure of the electron and further does not provide  answers for the physically important situation of continuous symmetries (like charge conservation). 
Finally even for bosonic systems it was not clear whether the group cohomology classification is complete. Indeed it is now known that in $3d$ there are SPT states that are not 
captured\cite{avts12,hmodl,burnellbc,cobord} by this classification. 

Tremendous progress has been made on these questions through a variety of less formal physics-based approaches in both $d = 2$ and $d = 3$. For bosonic systems they give an {\em understanding} of the phases predicted by the classification for many simple protecting symmetries, and predict further new phases.  Most crucially  physics-based methods  enable addressing electronic systems with realistic symmetries in the important case of three dimensional systems.   For instance for spin-orbit coupled electronic insulators in $3d$, 
it has been shown that there are precisely $6$ new topological insulating phases\cite{wpssc14} that have no non-interacting counterpart. Possible experimental signatures of these phases have been identified. These states have simple descriptions. They are either electronic Mott insulators where the spins have formed time reversal protected SPT states (as described in Refs. \cite{avts12,hmodl,burnellbc}) or are combinations of them and the conventional topological band insulator.  For topological insulators/superconductors with many other physically relevant symmetries\cite{TScSTO,wangts14} both the stability to interactions and the possible new phases have been determined. 

Why are these generalizations of the free fermion topological phases to interacting SPT phases interesting? First, because they may be there. In the context of ongoing experimental explorations of  topological phenomena in correlated quantum materials it is important to be aware of the possible interesting phases that might exist. Second, being short-range entangled, SPT phases provide what may be the simplest setting to study the interplay of three fundamental themes of quantum condensed matter physics: symmetry, strong correlation, and topology. These are the same ingredients in other frontier problems many of which involve long range entangled quantum ground states. 
Examples are quantum spin liquid phases of frustrated magnets\cite{balentsrev}, non-Fermi liquid metals\cite{vojta}, and quantum critical points beyond the Landau paradigm\cite{deccp}.  Studies of the relatively simple SPT phases have provided fresh and powerful theoretical insights into the physics of these complex long range entangled states, in particular on the realization of symmetry in these phases.

We will review the key ideas behind these developments here using several examples. The emphasis will be on providing physical intuition and insight with a focus on electronic systems in $3d$.  It is beyond the scope of this review however to describe group cohomology and other pertinent mathematical structures. A useful earlier review with more details on $1d$ and $2d$ systems is in Ref. \onlinecite{atav13}.

\section{Review of topological band insulators}
The three dimensional Topological Band Insulator (TBI) is known\cite{qi,essin} to be stable to weak  electron-electron interactions. 
Let us first review its physics  emphasizing properties that are robust in the presence of interactions. 
The TBI phase occurs in spin-orbit coupled electronic systems with charge conservation (corresponding to $U(1)$ phase rotations) and time reversal (denoted $Z_2^T$) symmetries.  If $c_\alpha$ represents the electron destruction operator of spin $\alpha$, 
these symmetries are implemented through 
\begin{eqnarray}
U^{-1} c_\alpha U & = & e^{i\theta} c_\alpha \\
{\cal T}^{-1} c_\alpha {\cal T} & = & i(\sigma^y)_{\alpha \beta} c_\beta
\end{eqnarray}
Note that time reversal is anti-unitary, and that the $U(1)$ phase rotation does not commute with the time reversal operation. Formally these define the symmetry group $U(1) \rtimes Z_2^T$.  The electron transforms as a Kramers doublet under time reversal.  

Within one electron band theory insulators with these symmetries come in two distinct classes: the conventional band insulator and the topological band insulator. A striking physical characterization of the TBI is in terms of its protected surface states. The surface is gapless with a Fermi surface that encloses an odd number of Dirac cones. A strictly two dimensional metal with these symmetries is prohibited from having such a Fermi surface. Thus the surface of the three dimensional TBI realizes symmetries in a manner forbidden in strictly two dimensional systems. 

Though the TBI is often described within band theory, its surface states have the remarkable property that they cannot be localized by disorder (at least in the non-interacting limit - see Section \ref{etisto}) so long as the symmetries of charge conservation and time reversal are retained. This is again in sharp contrast to a strictly two dimensional metal even with spin orbit coupling. Thus the TBI phase itself remains distinct from the conventional band insulator even in the presence of disorder.  

It is interesting to consider the fate of the surface when time reversal is broken by application of a magnetic field or by depositing a ferromagnetic thin film on top. The surface can then be gapped out at the expense of introducing a surface quantum Hall response (in units where the electron charge and Planck's constant $h$ are equal to $1$)
\begin{equation}
\sigma_{xy} = n + \frac{1}{2} 
\end{equation}
where $n$ is an integer.  The shift by $\frac{1}{2}$ from an integer distinguishes the surface from a strictly two dimensional systems of non-interacting electrons. A very useful way of thinking about this surface quantum Hall state is to consider a domain wall in the deposited ferromagnet across which the magnetization changes sign. The surface of the TBI induces a gapless chiral `edge' mode at this domain wall that is identical to the edge mode of the two dimensional integer quantum Hall effect. 

The  magneto-electric response described above is nicely encapsulated in a different way by considering the response of the bulk insulator to external electromagnetic fields.    For any insulator in 3D, the effective long wavelength Lagrangian for an external electromagnetic field obtained by integrating out all the matter fields will take the form 
\begin{equation}
{\cal L}_{eff} = {\cal L}_{Max} + {\cal L}_\theta
\end{equation}
The first term is the usual Maxwell term and the second is the `theta' term:
\begin{equation}
{\cal L}_\theta = \frac{\theta}{4\pi^2} \v{E}\cdot\v{B}
\end{equation}
where $\v{E}$ and $\v{B}$ are the external electric and magnetic fields respectively. 

Under time reversal, $\theta \rightarrow - \theta$ and in a fermionic system the physics is periodic under $\theta \rightarrow \theta + 2\pi$.  Time reversal symmetric insulators thus have $\theta = n \pi$ with $n$ an integer. Trivial time-reversal symmetric insulators have $\theta = 0$ while free fermion topological insulators have $\theta = \pi$\cite{qi,essin}.

A physical understanding of the periodicity is obtained as follows. If we allow for a boundary to the vacuum and further assume that the boundary is gapped, then the $\theta$ term leads to a surface Hall conductivity of $\frac{\theta}{2\pi}$. 
To see this, assume a boundary (say at $z = 0$), $\theta = \theta(z)$ is zero for $z< 0$ and constant $\theta$ for $z >  0$. The action associated with the $\theta$ term
is
\begin{eqnarray}
S_\theta & = & \frac{1}{8\pi^2} \int d^3x\, dt \,\theta(z) \partial_\mu K^\mu \\
& = & -\frac{1}{8\pi^2} \int d^3x\,dt \, \frac{d\theta}{dz} K^z \\
& = & \frac{\theta}{8\pi^2} \int_{\partial B} d^2x\,dt \epsilon^{z\nu \lambda \kappa} A_\nu \partial_\lambda A_\kappa
\end{eqnarray}
where $A$ is the external electromagnetic potential and $K^\mu = \epsilon^{\mu\nu\lambda\kappa} A_\nu \partial_\lambda A_\kappa$. This is a surface Chern-Simons term and leads to a Hall conductivity $\theta/2\pi$.

For fermion topological insulators $\theta = \pi$ so that the surface $\sigma_{xy} = \frac{1}{2}$. If we shift $\theta \rightarrow \pi + 2n\pi$, then the surface $\sigma_{xy} = (n + \frac{1}{2})$. This corresponds to simply depositing an ordinary integer quantum Hall state of fermions at the surface of this insulator -  hence this should not be regarded as a distinct bulk state so that the only non-trivial possibility is $\theta =  \pi$.

A very powerful theoretical device - which we will use later - is to  imagine introducing an external magnetic monopole as a source of the magnetic field. The $\theta$ term in the induced action implies that such a monopole carries electric charge\cite{witten} $\frac{\theta}{2\pi}$. In the TBI phase it follows that the monopole carries charge $1/2$. This fractional charge on an external  monopole provides an alternate characterization of the TBI phase that is equivalent to the characterization in terms of a $\theta$ term.

\section{Warm-up: Bosonic SPT phases in $d =2$}
Before launching into electronic SPT phases let us first study SPT phases in systems of interacting bosons. As usual in strong correlation problems bosons are expected to be a lot easier to handle than fermions. Further there are natural realizations of correlated bosons: indeed any quantum magnet may be fruitfully viewed as a strongly correlated bosonic system. Needless to say cold atomic gases provide another realization.

As already mentioned the Haldane spin-$1$ antiferromagnetic chain in $d = 1$ is the classic example of a $1d$ bosonic SPT phase. 
A path to understand the physics of $d = 2$ bosonic SPT phases (first deduced through the formal classification\cite{chencoho2011})  for a number of simple symmetries was pioneered in  Ref. \cite{levinstern} and particularly Ref. \cite{luav2012}  using a Chern-Simons/edge theory approach\footnote{This approach fails in some examples with more complex symmetries than the ones considered here.}.   Here  we will  review some examples of such phases with a view to gleaning physical insights that will be useful in understanding electronic SPT phases.  More details on these $2d$ boson SPT phases can be found in Ref. \cite{atav13}.

\subsection{d = 2: Integer quantum Hall effect for bosons}
A celebrated precursor to the electronic topological insulator is the integer quantum Hall state of electrons in two dimensions.  We now study the question\cite{chencoho2011,luav2012,tsml12} of whether interacting two dimensional bosons in a strong magnetic field can form an integer quantum Hall state without any exotic bulk excitations. There is a long history of study of quantum Hall states of bosons, including some with Hall conductivity quantized to be an integer. However till recently the states studied all had  intrinsic topological order in the bulk leading to excitations with anyonic self/mutual statistics. In the context of this review we are interested in an integer quantum Hall state of bosons which does not have such intrinsic topological order.  

The electronic integer quantum Hall effect is usually discussed in terms of completely filling a Landau level of states within an independent electron picture. In contrast for bosons, even the integer quantum Hall state requires interactions. 
 We now describe the simplest example of such a state using a physical model\cite{tsml12}.

Consider a system of two component bosons 
in a strong magnetic field  at a filling factor $\nu = 1$ for each boson species.  A natural realization is 
in terms of pseudospin-$1/2$ ``spinor" bosons of ultracold atoms in artificial gauge fields. 
Initially we assume that there is no inter-species tunneling but we can  relax this assumption 
later. The system then actually has $U(1) \times U(1)$ symmetry 
corresponding to separate conservation of the two species of bosons. The Hamiltonian is:
\begin{eqnarray}
H & = & \sum_I H_I + H_{int} \\
H_I & = & \int d^2x \sum_I b^\dagger_I \left(-\frac{\left(\vec \nabla - i\vec A \right)^2}{2m}
- \mu \right) b_I \\
H_{int} & = &  \int d^2 x \sum_I  g_s \left(\rho_I(x) \right)^2 + 2g_d  \rho_1(x) \rho_2(x) \nonumber
 \label{eq:hamiltonian}
\end{eqnarray}
Here $b_I$ is the boson annihilation operator for species $I$ where $I = 1,2$, and $\rho_I$ the corresponding density. The vector potential
$\vec A$ describes the external $B$-field.   When $g_s = g_d$ the Hamiltonian has an extra pseudospin $SU(2)$ symmetry which rotates the two species of bosons into one another.

\begin{figure}
 \includegraphics[scale=0.5]{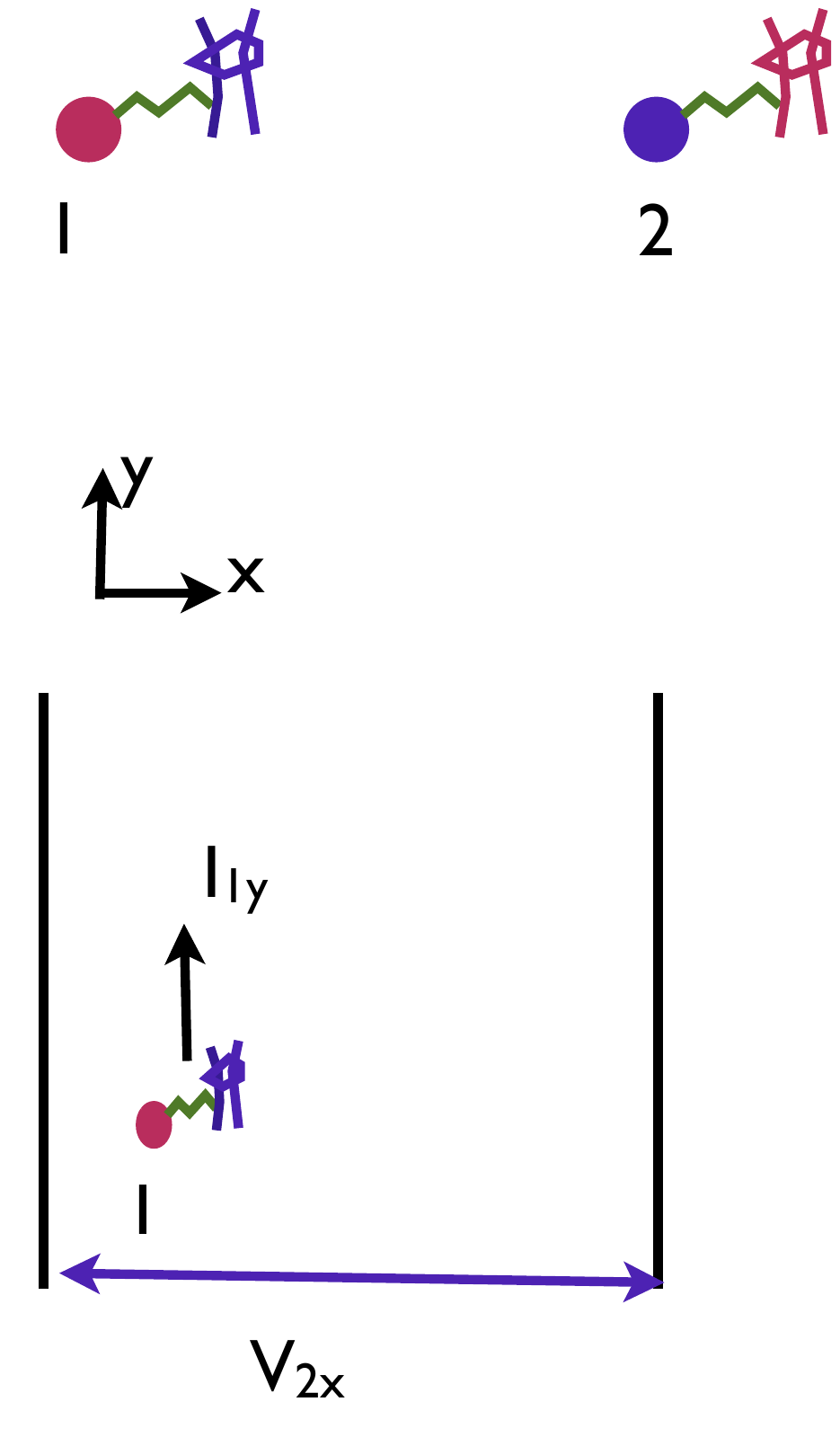}
\caption{Mutual composite bosons and their  transport in the 2-component boson integer quantum Hall effect }
\label{biqhetr}
\end{figure}

We construct a  candidate state using  a flux attachment Chern-Simons theory where we attach to each boson one flux quantum of the other species. 
Define new boson operators
\begin{eqnarray}
\tilde{b}_1(\v x) & = & e^{-i\int d^2 x' \Theta (\v x - \v x') \rho_{2}(\v x')} \cdot b_1(\v x) \\
\tilde{b}_2(\v x) & = & e^{-i\int d^2 x' \Theta (\v x - \v x')\rho_{1}(\v x')} \cdot b_2(\v x)
\end{eqnarray}
where $\Theta(\v x)$ is the angle at which the vector $\v x$ points.  We will call the bosons $\tilde{b}_{1,2}$ ``mutual composite bosons.''
With $\nu = 1$ for each species, we can clearly cancel the flux of the external magnetic
field in a flux smearing mean field approximation.   The mutual composite bosons can then condense and the result will be a quantum Hall state. Simple arguments determine the transport properties of this state. 
Consider passing currents $I_{1y}, I_{2y}$ of the two species along the $y$-direction (see Fig. \ref{biqhetr}). The $I_{1y}$ current corresponds to a flow of the mutual composite boson $\tilde{b}_1$. As this is attached to a flux quantum of species $2$, there is a voltage $V_{2x}$ induced along the $x$-direction (and similarly a voltage $V_{1x}$ induced by the current $I_{2y}$) given by 
\bea
V_{1x} & = & I_{2y} \\
V_{2x} & = & I_{1y}
\eea
(in units where $e = 2\pi \hbar = 1$). 
Now let us consider the total charge and pseudospin currents and the corresponding voltages: $I_{cy} = I_{1y} + I_{2y}, I_{sy} = I_{1y} - I_{2y}$, $V_{cx} = \frac{V_{1x} + V_{2x}}{2}, 
V_{sx} = \frac{V_{1x} - V_{2x}}{2}$. It follows that 
\bea
I_{cy} & = & 2V_{cx}\\
I_{sy} & = & - 2V_{sx}
\eea
Thus this state has electrical Hall conductivity $\sigma^c_{xy} = 2$ while the pseudospin Hall conductivity $\sigma^s_{xy} = -2$.  

We can try to construct excitations a 'la Laughlin. We thread in a $2\pi$ flux quantum at some point $z_0$ of the sample. This will pick up an electric charge $\sigma_{xy} = 2$.  Further if we exchange two such quasiparticles the phase we get is $\pi \sigma_{xy} = 2\pi$, {\em i.e} these excitations are bosons. Thus unlike for familiar quantum Hall states we do not get fractional charge or statistics through this construction. 

\begin{figure}
 \includegraphics[scale=0.5]{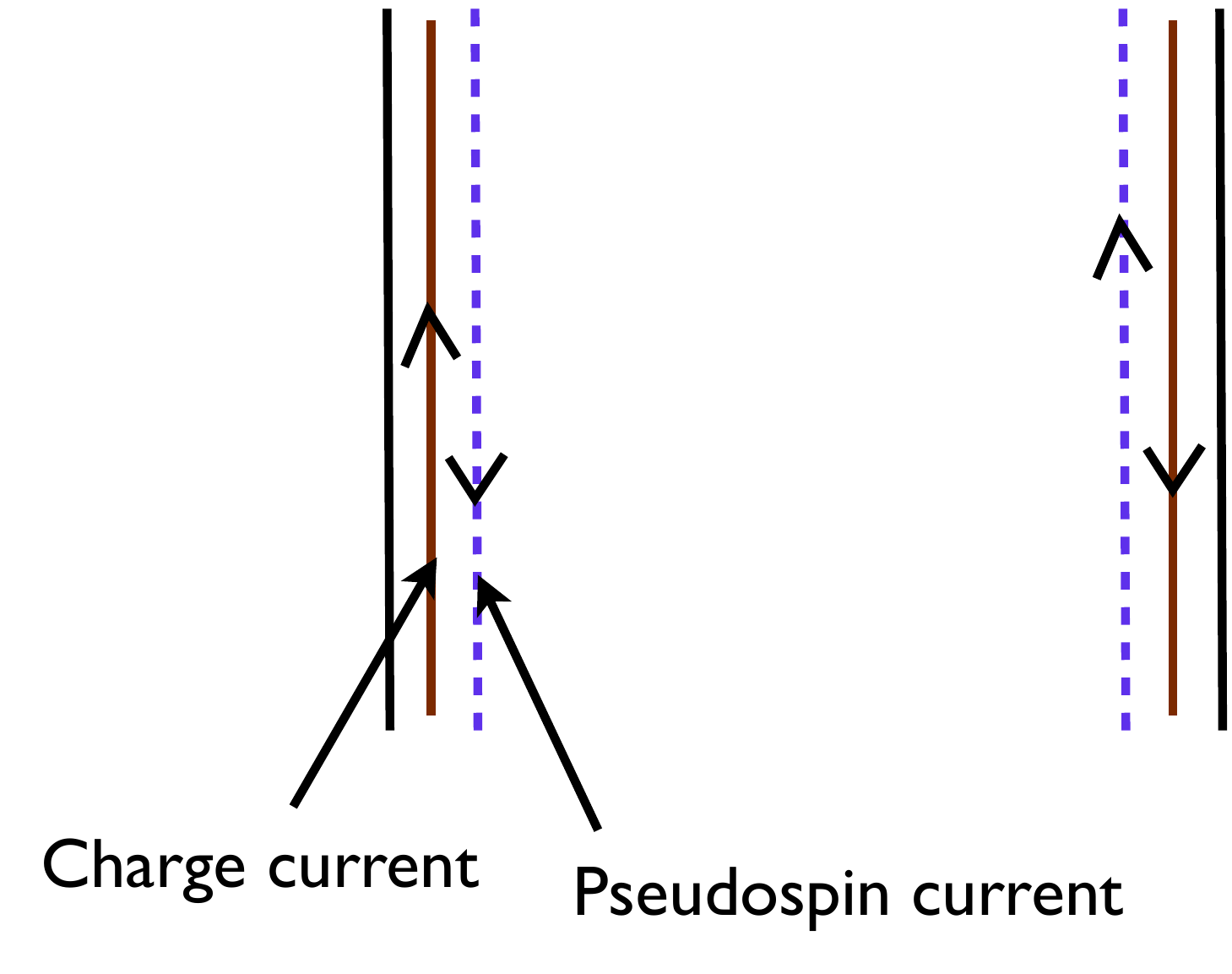}
\caption{Edge structure in the boson integer quantum Hall effect }
\label{biqheedge}
\end{figure}

Let us now describe the structure of the edge states. Consistent with the Hall conductivities, there are two counterpropagating chiral modes (see Fig. \ref{biqheedge}) of which one
carries electric charge, and the other is electrically neutral (but carries pseudospin).
As a result of this structure, the thermal Hall conductivity vanishes, even though
the electric Hall conductivity is nonzero. Note that backscattering between the two counterpropagating modes is prohibited by the
$U(1)$ charge conservation symmetry (even if we allow interspecies tunneling which breaks pseudospin conservation). Thus despite being non-chiral the gapless edge modes are
``symmetry-protected.''   

Even though the edge theory looks like a standard one dimensional theory, the realization of charge symmetry is `anamolous', and is forbidden in strictly one dimensional systems. This will be a recurring theme in the study of SPT phases. Indeed $2d$ SPT phases with many other symmetries have similar  `chiral' symmetry realization at the edge\cite{liuxgw}

This simple analysis is easily formalized through a Chern-Simons Landau-Ginzburg theory which can then be used to derive an effective Chern-Simons topological field theory in terms of two $U(1)$ gauge fields.  This formulation confirms that the state described is short range entangled with the properties described above.

Ref. \cite{tsml12} also proposed ground state wave functions for this state; for instance:
\begin{eqnarray}
\Psi_{flux} &=& P_{LLL} \prod_{i < j} |z_i - z_j|^2 \cdot \prod_{i < j} |w_i - w_j|^2 \nonumber \\
&\cdot& \prod_{i , j} (z_i - w_j) \cdot e^{- \sum_i \frac{|z_i|^2 + |w_i|^2}{4}}
\label{wf2}
\end{eqnarray}
where $P_{LLL}$ denotes the projection onto the lowest Landau level and $z_i, $ $w_i$ are the complex coordinates of the particles of two boson species respectively.  

This wave function is a spin singlet under the $SU(2)$ pseudospin symmetry\footnote{To see this
note that, before projection, it can be written as a product
of the anti-analytic spin-singlet $(221)$ state and a fully symmetric function of $z_i, w_j$.}, suggesting that the $SU(2)$ symmetric Hamiltonian $g_s = g_d$ may support this boson integer quantum Hall state. 
Finally several exact diagonalization studies\cite{ueda13,wujain13,nrts13} have been performed on the model and find strong evidence for the boson integer quantum Hall state.

Having  constructed an integer quantum Hall SPT state for bosons
with an electric Hall conductivity of $\sigma_{xy} = 2$ (in appropriate units), clearly we can obtain other states with $\sigma_{xy} = 2n$ by taking copies. The formal Chern-Simons classification\cite{luav2012} shows that $\sigma_{xy}$ cannot be odd for such states. To understand this simply\cite{tsml12} consider 
a general bosonic quantum Hall state, and imagine threading in $2\pi$ flux at some point  $z_0$. This produces an excitation with charge $\sigma_{xy}$, and braiding statistics $\pi \sigma_{xy}$. However in an SPT state of bosons all excitations must have bose statistics. It follows that $\sigma_{xy}$ must be even. 

Other realizations of the boson integer quantum Hall state have been studied. Ref. \cite{lesik13} constructed lattice models for this phase that admit a sign-problem free formulation and enabled  studying the edge through Monte Carlo simulations. Refs. \cite{lulee12,yewen13} described parton constructions for lattice system with $U(1)$ symmetry, and proposed candidate Hamiltonians. 

In general a quantum Hall phase also has a quantized thermal Hall conductivity $\kappa_{xy} = \nu_Q \frac{\pi^2}{3}\frac{k_B^2}{h} T$ where $k_B, T$ are Boltzmann's constant and the temperature respectively.  The constant $\nu_Q$ counts the different between the number of counter propagating edge modes 
(sometimes known as the chiral central charge), and is a universal topological property of the state.  The states we have discussed so far have $\nu_Q = 0$. A different class of un-fractionalized bosonic state that requires no symmetry at all is possible so long as $\nu_Q$ is a multiple of $8$. These have been discussed in Refs. \cite{kitaev06,luav2012,nayake8,sss14}.

\subsection{Ising SPT in d = 2: Braiding statistics}
We now discuss another 2d SPT phase, now in spin systems with Ising ({\em i.e} $Z_2$ symmetry).  The familiar transverse field Ising model in $d = 2$ space dimensions has both an ordered phase and a paramagnetic phase. This paramagnetic phase is smoothly connected to a product state and we will call it the trivial paramagnet. As shown in Refs. \cite{xie2dti,levingu} - though it is realized only in more complicated models - quantum magnets with Ising symmetry support an SPT phase. We will call this phase a `topological Ising paramagnet'. 

How is the topological Ising paramagnet different from a  trivial Ising paramagnet? As an SPT phase it has protected edge states: the edge is either gapless or spontaneously breaks the Ising symmetry. An insightful discussion of the fundamental physics of the topological Ising paramagnet was provided in Ref. \cite{levingu}.  Consider the ground state wave function of Ising paramagnets. It is useful to think of the paramagnetic state as arising out of the ordered Ising ferromagnet by proliferating domain walls of the Ising order. A prototypical wave function for the trivial Ising paramagnet is simply an equal amplitude sum over all domain wall loops:
\be
|P\> =\sum_{D} | D \> 
\ee
For the  topological Ising paramagnet on the other hand a prototypical wave function is
\be 
|TP\> =\sum_{D} \left(-1\right)^{N(D)}| D \> 
\ee
Here $N(D)$ is the total number of loops in the domain wall configuration $D$. The key difference is the extra phase factor that weights configurations with an odd number of loops with a relative $-$ sign.  Following the discussion on TBI phases of electrons in $3d$, we can formally imagine gauging the $Z_2$ symmetry, and studying the fate of the resulting $Z_2$ fluxes. In the trivial paramagnet these fluxes (which are point particles) have either bose or fermi statistics (these two possibilities differ by attaching an Ising spin, {\em i.e} $Z_2$ charge to the flux). In contrast in the topological paramagnet the extra $\left(-1\right)^{N(D)}$ phase factor leads to semion or antisemion statistics\cite{levingu} for the $Z_2$ flux (again the 2 possibilities differ by attachment of Ising spin). Thus the braiding statistics of fluxes obtained by gauging the global symmetry is a diagnostic of the SPT order in the phase. 

These features of the ground state wave function and in the statistics of the fluxes can be shown to directly lead to the symmetry protected edge states of the topological Ising paramagnet\cite{levingu}.

\section{Bosonic Topological Insulators in d = 3}
\label{bti3d}
We now turn to the physics of time reversal invariant bosonic topological insulators in 3d. We will consider two distinct physical situations which have both boson number conservation and time reversal symmetries. For bosons (such as He-4) the $U(1)$ phase rotation corresponding to boson number conservation does not commute with time reversal. The corresponding symmetry group is $U(1) \rtimes Z_2^T$. A different situation arises in quantum spin models with time reversal symmetry with, in addition, XY symmetry corresponding to conservation of $S_z$. Then the $U(1)$ group of spin rotations about the $z$ axis commutes with the time reversal transformation and the corresponding symmetry group is $U(1) \times Z_2^T$. 

We begin by constraining the EM response of a bosonic topological insulator when the symmetry group is $U(1) \rtimes Z_2^T$.  
Again T-reversal and periodicity imply $\theta = n\pi$ and a surface $\sigma_{xy} = n/ 2$. A crucial observation\cite{avts12} is that now $\theta = 2\pi$ must be regarded as {\bf distinct} from $\theta = 0$. At $\theta = 2\pi$ the surface $\sigma_{xy} = 1$. But this cannot be obtained from the surface of the $\theta = 0$ insulator by depositing any 2d integer quantum Hall state of bosons.  Thus the surface state of the $\theta = 2\pi$ boson insulator is not a trivial 2d state but rather requires the presence of the 3d bulk.
Therefore $\theta = 2\pi$ necessarily corresponds to a non-trivial 3d bosonic TI\cite{avts12}. 
 
Exactly the same result also describes bosons with symmetry $U(1) \times Z_2^T$ appropriate for spin systems: the response to external gauge fields that couple to the conserved global $U(1)$ current has a $\theta$ term where $\theta = 2\pi$ is a distinct state from $\theta = 0$. With these symmetries however we will see below that even at $\theta = 0$ there are two {\em distinct} topological insulators\cite{avts12}. Thus the $\theta$ value in the EM response does not uniquely define the SPT order\footnote{More precisely we are referring to SPT phases protected by the full symmetry. For both symmetries in addition to the phases discussed in this section there are SPT phases which are protected just by the $Z_2^T$ alone. These additional states will be discussed in Section \ref{tp}. Including these the full classification is $Z_2^3$ for bosons with $U(1) \rtimes Z_2^T$, and $Z_2^4$ for bosons with $U(1) \times Z_2^T$ symmetry. }.  The different SPT phases for either symmetries are summarized in Table. \ref{btitable}. 

 \begin{table*}[tttt]
\begin{tabular}{|>{\centering\arraybackslash}m{1.5in}|>{\centering\arraybackslash}m{1.2in}|>{\centering\arraybackslash}m{1in}|>{\centering\arraybackslash}m{1in}|}
\hline
{\bf Symmetry } &  {\bf Label} & {\bf $\theta$ value} & {\bf Surface vortex/bulk monopole} \\ \hline
 
 $U(1)\rtimes\Z_2^T$   & $F$ & $2\pi$ & fermion \\ \hline
 \hline
$U(1)\times\Z_2^T$   & $ F$ & $2\pi$ & fermion, Kramers singlet\\ \hline
 
 \hline
$U(1)\times\Z_2^T$ & $ K$ & $0$ &  Kramers doublet  \\ \hline
 $U(1)\times\Z_2^T$ &  $FK$ & $2\pi$ & fermion, Kramers doublet \\ \hline

\end{tabular}
\caption{Some properties of the various boson topological insulators in $3d$. There are additional SPT phases protected by $Z_2^T$ alone which are discussed in Section \ref{tp}.
The labels describe the surface vortex or equivalently the bulk neutral monopole as described in the text. }
\label{btitable}
\end{table*}%

\subsection{Surface theory}
We now discuss the surface theory of these bosonic topological insulators. A key physical requirement  on any putative effective theory of the surface is that it does not admit a trivial insulating phase.  Implementing this leads very directly to a powerful effective `Landau-Ginzburg' theory of the surface\cite{avts12}. 

Let us consider a superfluid state at the surface that spontaneously breaks the global $U(1)$ symmetry. It will be extremely useful to formulate the surface effective theory in terms of vortices of the superfluid order parameter. Indeed for bosons in two dimensional systems there exists a duality transformation\cite{mp78,cdbh81,fl89} that trades a formulation in terms of the physical charge-carrying bosons for a different formulation in terms of vortex degrees of freedom.  Specifically there is a dual effective Landau-Ginzburg theory with the Lagrangian written schematically as
\begin{equation}
\label{bvLG}
{\cal L}_d  = {\cal L}[\Phi_v, a_\mu] + \frac{1}{2\pi}A_\mu \epsilon_{\mu\nu\lambda} \partial_\nu a_\lambda
\end{equation}
The first term describes a bosonic field $\Phi_v$ coupled minimally to a dual internal gauge field $a_\mu$, and $A_\mu$ is an external probe gauge field. The field $\Phi_v$ describes the vortex of the original physical boson. The global $U(1)$ current is 
\begin{equation}
j_\mu = \frac{1}{2\pi}  \epsilon_{\mu\nu\lambda} \partial_\nu a_\lambda
\end{equation}

In the superfluid phase the vortices are gapped. Consequently in the  dual formulation the superfluid is understood as a Mott insulator of the vortices\cite{fl89}. A  trivial boson insulator is, on the other hand,  obtained by condensing the  $2\pi$ vortices\cite{fl89} described by $\Phi_v$.  This quantizes the flux of the dual vector potential $\int  d^2x (\partial_x a_y - \partial_y a_x)$ to multiples of $2\pi$ which corresponds precisely to the quantization of particle number expected in the boson insulator.

Turning now to the surface of the three dimensional bosonic topological insulator, it is clear that to implement the absence of a trivial insulator  we must require that these $2\pi$ vortices not be able to condense.  This is very simply guaranteed if the vortex is a fermion\cite{avts12} rather than a boson! 

Let us now explore this idea seriously. 
With a fermionic vortex field which we denote $c$, the surface Landau-Ginzburg Lagrangian may be written schematically 
\begin{equation}
\label{fvLG}
{\cal L}_d = {\cal L}[c, a_\mu] + \frac{1}{2\pi}A_\mu \epsilon_{\mu\nu\lambda} \partial_\nu a_\lambda
\end{equation}
The important modification from the standard dual Lagrangian (Eqn. \ref{bvLG}) described above   is in the statistics of the vortex field which precludes $2\pi$ vortices from condensing. 
 As usual if the $c$-fermionic vortex is gapped (say in an ordinary band insulator) then we get the surface superfluid phase. 
 As described in Ref. \cite{avts12,hmodl}  if we break time reversal at the surface we can get a gapped phase without topological order. This is obtained by simply letting the $c$-fermionic vortices completely fill a topological band with Chern number $\pm 1$, {\em i.e} the Hall conductivity of the $c$-fermion is $\sigma^c_{xy} = \pm 1$.  It is readily seen that the surface then has an electrical Hall conductivity $\sigma_{xy} = \pm 1$  
 exactly consistent with a bulk $\theta$ term in the electromagnetic response with $\theta = 2\pi$. Thus the dual Landau-Ginzburg theory correctly describes a $\theta = 2\pi$ boson  topological insulator. As the surface vortex is a fermion we will label this phase $F$. 

If we were to discuss SPT phases of time reversal symmetric spin systems with conserved $S^z$ (symmetry $U(1) \times Z_2^T$) then we can again start with the $XY$ ordered phase and create an obstruction for $2\pi$ vortices of the $XY$ spin order to condense. But now there is more than one way to create this obstruction\cite{avts12} corresponding to more than one non-trivial topological insulating phase. The key point is that with this symmetry the vorticity is even under time reversal so that a vortex stays a vortex. Then it makes sense to ask if the vortex is Kramers or not under time reversal ({\em i.e} if $T^2 = \pm 1$). Note that the vortices are non-local objects and hence are allowed to transform projectively ({\em i.e} as Kramers) under time reversal.  In contrast for bosons with $U(1) \rtimes Z_2^T$ the vortex goes to an anti vortex under time reversal, and the question of Kramers or not does not arise. Thus for spin systems, the obstruction to condensing $2\pi$ vortices can be implemented by choosing them to be (1) bosonic Kramers doublets (we label this phase $K$) (2) fermionic Kramers singlet (label $F$) or (3) fermionic Kramers doublets (label $FK$). In the first case the vortices can condense but the condensate necessarily breaks time reversal. These different possibilities correspond to distinct bulk SPT phases for spin systems with symmetry $U(1) \times Z_2^T$.

With a bosonic Kramers doublet vortex $z_\alpha$, the dual Landau-Ginzburg theory takes the form
\begin{equation}
{\cal L}_d = {\cal L}[z_\alpha, a_\mu] + \frac{1}{2\pi} A da
\end{equation}
Under time reversal $z_\alpha \rightarrow i\sigma^y_{\alpha\beta} z_\beta$.   It is readily seen in this case that if we break time reversal to produce a surface state without topological order then it has $\sigma_{xy} = \kappa_{xy} = 0$. Thus the corresponding bulk SPT state responds to external gauge fields coupling to the $U(1)$ currents with $\theta = 0$.

The surface phase structure of these other SPTs can be readily discussed in terms of these dual Landau-Ginzburg theories\cite{avts12,hmodl}.

\subsection{Monopoles in the bulk}
For the 3d electronic TBI phase, we saw that the $\theta$ term in the bulk EM response implies that external magnetic monopoles carry a fractional charge $1/2$. 
This fractional charge on the monopole provides an interesting characterization of the TBI phase. 
Similarly for the 3d bosonic topological insulators (with either $U(1) \rtimes Z_2^T$ or $U(1) \times Z_2^T$ symmetry) we can ask about the structure of the magnetic monopole 
when the global $U(1)$ currents are coupled to external gauge fields. 

In the bosonic case the properties of the bulk monopole are directly inherited from the properties of the surface vortex in the dual Landau-Ginzburg description. Indeed if we gauge the global $U(1)$ symmetry and imagine tunneling a monopole from the vacuum into the bulk it will leave behind at the surface precisely the $2\pi$ vortex. Since the monopole in the vacuum is a trivial boson the exotic vortex left behind determines the properties of the monopole in the bulk of the SPT phase. Thus for the bosonic TI with $U(1) \rtimes Z_2^T$ symmetry we conclude that there is a bulk monopole that is electrically neutral and has fermionic statistics (dubbed the `statistical Witten effect'\cite{statwitt}). For the bosonic TIs with $U(1) \times Z_2^T$ symmetry for the three non-trivial SPT phases discussed above, the bulk monopole is electrically neutral, and is either (i) a Kramers doublet boson (which we dub the `Kramers Witten effect') or (ii)  a Kramers singlet fermion or (iii) a Kramers doublet fermion. 

The monopole structure for bosonic TIs with a $\theta = 2\pi$ EM response can also be determined directly\cite{statwitt} through a different argument which we illustrate for  $U(1) \rtimes Z_2^T$ symmetry. Consider first the theory when $\theta = 0$ and imagine gradually increasing $\theta$.  At $\theta = 0$ the monopole is a trivial electrically neutral boson. For general $\theta$, the monopole carries electric charge $\frac{\theta}{2\pi}$. So when $\theta = 2\pi$, it carries charge-$1$. But now we can obtain a neutral monopole by binding the elementary charge $-1$ boson to this monopole. However as is well known binding unit charge to  a unit monopole changes the statistics from boson to fermion. We thus conclude that the neutral monopole is a fermion. 

This argument may be used as an alternate logical starting point to discuss the physics of the bosonic TI with $U(1) \rtimes Z_2^T$ symmetry.   As elegantly argued in Ref. \cite{statwitt}, the non-trivial structure of the bulk monopole can be used to constrain the surface physics.  We will illustrate this in the next subsection.

Finally a physical picture\cite{avts12,xuts13} of the bulk of the various boson SPT phases we have discussed is very useful to keep in mind. In general any 3d boson insulator can be regarded as descending from a 3d superfluid by proliferating vortex loops.  The different boson SPTs are distinguished by the structure of these vortex loops. One possibility is that the vortex loops should be viewed as ribbons and the vortex loop condensate has a phase $(-1)$ associated with the self-linking of these ribbons\cite{xuts13}.  At the interface with the vacuum (or a trivial boson insulator) these vortex ribbons terminate into point vortices. The self-linking phase in the bulk directly then leads to fermi statistics of these surface vortices\cite{xuts13} corresponding to the $F$ phases in Table. \ref{btitable}. Similarly  if the global $U(1)$ symmetry is gauged we identify the vortex lines with $2\pi$ magnetic flux lines; open ends of these lines are the magnetic monopoles. These again will have Fermi statistics coming from the self-linking phase.  For  the $K$ state (allowed for symmetry $U(1) \times Z_2^T$) a different option is possible. The vortex loop may have a time reversal symmetric Haldane chain residing in its core\cite{avts12,xuts13}. As the Haldane chain is gapped naively there seems to be nothing special about such a vortex loop. However at the surface the points of penetration of the vortex line expose an open end of the Haldane chain. It follows that these are Kramers doublet surface vortices. Similarly bulk monopoles of external gauge fields will correspond to open ends of bulk vortex lines which will therefore also be Kramers doublets. 

A complementary physical picture is to view the $F$ state (for $U(1) \times Z_2^T$ symmetry) as descending from a state with broken $Z_2^T$ symmetry by proliferating domain walls. If these domain walls are decorated\cite{clav14} with the $2d$ boson integer quantum Hall state then a $3d$ SPT phase results. 

\subsection{Surface topological order}
A very interesting possible surface phase is one that preserves all symmetries and is gapped. The price is that such a surface must have intrinsic topological order\cite{avts12,swingle} ({\em i.e} must be long range entangled even though the bulk is not). A key property of this surface topological order is that the physical symmetries are implemented in an `anamolous' way, {\em i.e}, in a manner prohibited in strictly two dimensional systems with the same topological order\cite{avts12}.  Surface topological ordered phases have played a crucial conceptual role in our understanding of interacting 3d SPT phases. 

Let us illustrate the physics of the surface topological ordered phase in the example of the boson topological insulator. As discussed above the surface is described by the dual Landau-Ginzburg theory of Eqn. \ref{fvLG} with a fermionic vortex. Though this prevents single vortices from condensing to produce a trivial insulator, 
surface superfluid order can be killed if pairs of $c$-vortices condense, 
 {\em i.e} $<cc> \neq 0$.  As is familiar from discussions of fractionalized boson insulators in two dimensions\cite{bfn,z2long}, this leads to a surface topological order described by a deconfined $Z_2$ gauge theory.  
 The unpaired $c$-fermion survives as a gapped excitation carrying zero global $U(1)$ charge.  We call this the neutral $\epsilon$ particle. The pair condensation quantizes flux of $a_\mu$ in units of $\pi$. The result is a bosonic particle that  carries global $U(1)$ charge $1/2$ that we dub the $e$ particle.  Note that the $e$ and $\epsilon$ are mutual semions
 as expected for a deconfined $Z_2$ gauge theory. Their bound state - denoted the $m$ particle - also has charge-$1/2$ and is a mutual semion with both $e$ and $\epsilon$.  As both $e$ and $m$ carry fractional charge we denote this phase the $eCmC$ $Z_2$ topological  order\cite{avts12,hmodl}. 
 
 The topological content of this gapped symmetry preserving surface state is not at all unusual.  What is unusual is the symmetry realization: The charge assignment is inconsistent with time reversal\cite{avts12} in any strictly $2d$ system even though it naturally emerges at the surface of the 3d boson TI.  To see this simply\cite{statwitt}, let us imagine 
 threading $2\pi$ flux through the system.  If we think of the $2d$ system as embedded in $3d$ space we can think of this as tunneling a  magnetic monopole through the $2d$ sample.  As the electrical Hall conductivity is zero (by time reversal) this flux insertion acquires no charge and must create a bosonic excitation.  Equivalently since the magnetic monopole in the vacuum is a neutral boson its tunneling through the $2d$ sample will also leave behind a neutral boson. Now in the $eCmC$ phase this neutral excitation is seen by both the $e$ and $m$ particles as $\pi$ flux  (due to their charge of $1/2$).  We therefore identify the relic of the flux threading with the $\epsilon$ particle. However this is a fermion contrary to what we inferred above.  Thus the $eCmC$ state is forbidden in strict $2d$.  At the surface of the $3d$ boson TI however the monopole in the bulk is a fermion. Consequently when we tunnel in the monopole from the vacuum its relic at the surface will also be a fermion, and the $eCmC$ state becomes allowed.

\subsection{Coupled layer construction}
It should now be clear that to construct a 3+1-D SPT state, we only need to construct the corresponding topological order on the surface but have a confined bulk with gapped excitations. An  explicit such construction was described in Ref. \cite{hmodl} in  a $3d$ system built out of coupled layers of  $2d$ systems which we now illustrate by constructing the $eCmC$ state with $U(1) \rtimes Z_2^T$ symmetry.  

\begin{figure}
 \includegraphics[scale=0.5]{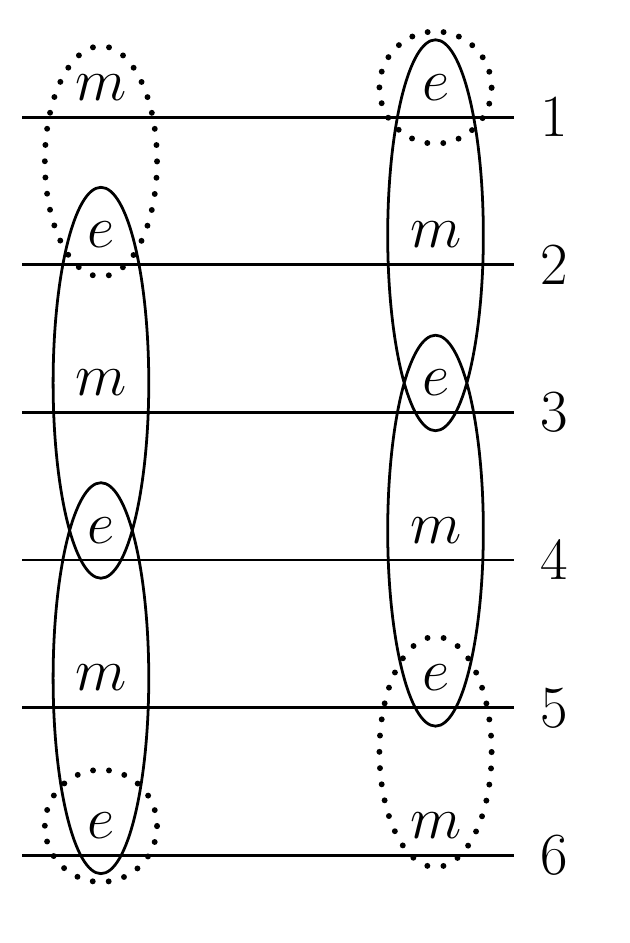}
\caption{Coupled-layer construction of SPT states. The particle composite in the ellipses are condensed, and only the four surface particles 
in the dotted ellipses survived as deconfined topological quasi-particles.}
\label{layer}
\end{figure}

In each layer consider a boson system in a fractionalized insulator with $Z_2$ topological order such that the $e$ particle carries charge-$1/2$ while the $m$ particle is neutral. Further assume that under time reversal both $e$ and $m$ are invariant. Such states are {\em allowed} in strictly $2d$ (for microscopic models see Refs. \cite{bosfrc,bosfrc3d}). 
Consider stacking $N$ layers of such states.  Now turn on an inter-layer coupling to make the composite particles $e_i^\dagger m_{i+1}e_{i+2}$ condensed, where $i$ is the layer index running from 
$1$ to $N-2$.  Note that the $e_i^\dagger m_{i+1}e_{i+2}$ all have bosonic self and bosonic mutual statistics so that they may be simultaneously condensed. As illustrated in Figure \ref{layer}, this procedure confines all the non-trivial quasi-particles in the bulk but not at the surface. For instance at the top surface    $e_1$ and $m_1e_2$ survive as deconfined excitations which are mutual semions and have self-boson statistics. Thus they form a $Z_2$ topological order at the top surface.  However  they both have global $U(1)$ charge $1/2$. so that the  surface is in the 
$eCmC$ state though the bulk has no exotic excitations.  By the analysis above we identify this with the 3d   bosonic TI with $U(1) \rtimes Z_2^T$ symmetry. 

The coupled layer construction is readily generalized to the other 3d boson SPT states discussed in this section and to  others discussed in Section \ref{tp}. 

\section{Electronic SPT phases in 1d and 2d}
Spin-orbit coupled insulators in $2d$ admit a TBI phase (corresponding to a $Z_2$ classification). With interactions the Chern-Simons/edge theory approach may be used to show that this remains unmodified by the presence of interactions. The 2d TBI is stable, and no further new phases are introduced(see upcoming arxiv version of \cite{luav2012}).  In $d = 1$
there is no TI phase even within band theory, and this remains true in the interacting system\cite{rosch,1dsptclass,1dfSPT}.  In particular the Haldane phase of spin systems - which is an SPT phase protected by just the $Z_2^T$ symmetry - is adiabatically connected to a trivial band insulator in an electronic system\cite{rosch}. Thus in $d = 1$ and $2$, interactions do not change the band theory classification of spin-orbit coupled insulators. We will see below that  $3d$ insulators are different and interactions induce additional phases not present within band theory.  Insulators with other symmetries in $2d$ have been discussed in Refs. \cite{luav2012,ncmt14}. 

For topological superconductors it was first shown in Ref. \cite{1dfSPT} that in $d = 1$ there are symmetries where interactions destabilize phases. Specifically a $Z$ classification of certain topological band superconductors was shown to collapse to $Z_8$ in the presence of interactions. A similar phenomenon was later demonstrated for some $2d$ topological superconductors\cite{2dfSPT,ncmt14}. We will discuss $3d$ topological superconductors below in Section \ref{tsc3d}.

\section{Electronic topological insulators in 3d}
\label{eti}

\subsection{Classification}
We are now ready to discuss interacting electronic topological insulators  (SPT) in $3d$.   
We focus initially on spin-orbit coupled insulators with the same symmetries  that   protect the familiar electronic topological band insulator, namely time reversal and charge conservation.  We review recent progress\cite{wpssc14} showing that with apart from the topological band insulator, there are 6 other non-trivial topological insulating states that {\em require} the presence of interactions. The appropriate  generalization of the $Z_2$ classification of band insulators is classification by the group $\mathbb{Z}_2^3$. This group structure means that all these interacting topological insulators can be obtained from 3 `root' states and taking combinations.  One of these root states is the standard topological band insulator. The other two require interactions and can be understood as Mott insulating states of the electrons where the resulting quantum spins have themselves formed a time reversal protected SPT phase. Such SPT phases of quantum spins (dubbed `topological paramagnets') were described in  Ref. \cite{avts12}. The three root states and their properties are  briefly described in Table. \ref{root}.

 \begin{table}[tttt]
\begin{tabular}{|>{\centering\arraybackslash}m{0.9in}|>{\centering\arraybackslash}m{1.5in}|>{\centering\arraybackslash}m{1in}|>{\centering\arraybackslash}m{1in}|}
\hline
{\bf Topological Insulator} &  {\bf Representative surface state} & {\bf $\mathcal{T}$-breaking transport signature} & {\bf $\mathcal{T}$-invariant  gapless superconductor} \\ \hline
Free fermion TI & Single Dirac cone & $\sigma_{xy}= \frac{\kappa_{xy}}{\kappa_0} =\pm1/2$ & None \\ \hline
Topological paramagnet I ($eTmT$) &  $\mathbb{Z}_2$ spin liquid with Kramers doublet spinon($e$) and vison($m$) & $\sigma_{xy} = \kappa_{xy} = 0$ & $N=8$ Majorana cones \\ \hline
Topological paramagnet II ($e_fm_f$) &  $\mathbb{Z}_2$ spin liquid with Fermionic spinon($e$) and vison($m$) & $\sigma_{xy} = 0; \frac{\kappa_{xy}}{\kappa_0}=\pm4$ & $N=8$ Majorana cones \\ \hline

\hline
\end{tabular}
\caption{Three root non-trivial topological insulators, with  representative symmetry-preserving surface states, and surface signatures when either time-reversal or charge conservation is broken on the surface (with topological orders confined). $\sigma_{xy}$ is the surface electrical Hall conductivity in units of $\frac{e^2}{h}$. $\kappa_{xy}$ is the surface thermal Hall conductivity  and $\kappa_0 =  \frac{\pi^2}{3}\frac{k_B^2}{h} T$ ($T$ is the temperature). $N$ is the number of gapless Majorana cones protected by time-reversal symmetry when the surface becomes a superconductor. A combination of these measurements could uniquely determine the TI. 
 }
\label{root}
\end{table}%

In general we may attempt to construct possible SPT phases of fermion system by first forming bosons as composites out of the fermions and putting the bosons in a bosonic SPT state.  However not all these boson SPTs remain distinct states in an electronic system.  The distinct ones can all be understood as electronic Mott insulators in spin-SPT states\cite{wpssc14}. 
Note also the contrast to the $1d$ and $2d$ cases where boson SPTs formed out of the electrons do not add any new phases beyond the band theory classification.

\vspace{4pt}
\noindent{\bf Generalities}:  Time reversal symmetric electronic SPT insulators have $\theta = n \pi$ with $n$ an integer. Trivial time-reversal symmetric insulators have $\theta = 0$ while free fermion topological insulators have $\theta = \pi$\cite{qi}. Suppose that for interacting electrons there is a new topological insulator that also has $\theta = \pi$. Then by combining it with the usual one we can produce a TI with $\theta = 0$. Thus without loss of generality we can restrict attention to the possibility of new TIs which have $\theta = 0$. 

Ref. \cite{wpssc14} employed the route of first constraining monopole sources of the external magnetic field, and then using these to constrain the surface.   
At $\theta = 0$, the elementary monopole carries zero electric charge.  Under time reversal the monopole becomes an anti-monopole as the magnetic field is odd. Together these fix the symmetry properties of the monopoles. In particular as the time reversed partner of a monopole lives in a different topological sector with opposite magnetic charge it is meaningless to ask if it is Kramers doublet or not\footnote{This should be contrasted with the discussion in Section \ref{bti3d}  on boson systems with $U(1) \times Z_2^T$ symmetry. There the monopole stays a monopole under time reversal and hence has the possibility of being a  Kramers doublet.}. 

There are still in principle two distinct choices corresponding to the statistics of the monopole: it may be either bosonic or fermionic.   
Ref. \cite{wpssc14} showed that bosonic monopoles only allow for the topological paramagnets mentioned above while fermionic monopoles  are forbiddden to occur in strictly three dimensional SPT systems built out of  charge-$1$ electrons.  Below we will describe the essential ideas involved, and refer the reader to the original paper for details.

\vspace{4pt}\noindent{\bf Topological insulators at ${\theta=0}$ - bosonic monopoles:} 
Following the discussion of bosonic topological insulators it will be very helpful for us to begin by thinking about a superconducting surface where the $U(1)$ symmetry is spontaneously broken. We then obtain symmetry preserving surface states by `quantum disordering' the superconductor.  To that end it is convenient to think in terms of a dual formulation\cite{z2long} of the surface superconductor in terms of vortices and other electrically neutral excitations (like the neutralized Bogoliubov quasiparticles)\footnote{To be general, we can allow the neutral sector itself to have topological order or even be gapless}..
Consider  any $3d$ insulator with  $\theta = 0$ and a bosonic monopole, and such a superconducting surface. 
 Now imagine tunneling a monopole from the vacuum to the system bulk. Since the monopole is trivial (chargeless and bosonic) in both regions, the tunneling event - which leaves a $2\pi$-vortex (or $\frac{hc}{e}$ vortex in the usual units) on the surface - also carries no non-trivial quantum number.  Furthermore this $\frac{hc}{e}$ vortex is local with respect to all the other excitations in the system. 
 
We emphasize that we have, at this stage, no constraints on the fundamental $\frac{hc}{2e}$ vortex. However since the $\frac{hc}{e}$ vortex is trivial we can condense it to obtain a symmetry preserving surface state.  As is well-known from descriptions of spin-charge separation in $2d$ the resulting state has distinct topological sectors. 
 The $\frac{hc}{e}$ condensate quantizes the electric charge in each topological sector to be an integer $q = ne$. 
 However, it is always possible to remove integer charge from the excitations in any  topological sector by binding physical electrons.  Thus  the theory of the surface factorizes into an electrically neutral sector (which may itself be non-trivial) supplemented with a gapped electron in a trivial surface insulator.  Furthermore the neutral sector is closed under time reversal. The only potentially non-trivial physics at the surface thus lives in this neutral sector.  But in an electronic system any local charge-neutral object has to be bosonic. 
 Hence the surface theory should be viewed as emerging purely from a neutral boson system. 
  
This implies that the bulk SPT order, if any,  should also be attributed to the neutral boson (spin) sector, {\em i.e}  it should be a time reversal protected SPT of spins in a Mott insulating phase of the electrons.  The SPT states of neutral bosons with time-reversal symmetry are classified\cite{avts12,hmodl,burnellbc} by $\mathbb{Z}_2^2$, with two fundamental root non-trivial phases. As we describe these below these can be characterized by their symmetry preserving surface topological orders.  Furthermore since even in the presence of electrons, the surface theory factorizes into this topologically ordered neutral sector supplemented with a gapped charge electron it follows that (unlike in $d = 1$) these spin-SPTs stay distinct from the trivial band insulator.  Combining this result with the $Z_2$ classification of band insulators, we arrive at the promised $Z_2^3$ classification of spin-orbit coupled electronic insulators in three dimensions. 

\vspace{4pt}\noindent{\bf Fermionic monopoles:} To complete the argument we need to dispose of the possibility that the monopole may be fermionic.  This was done in Ref. \cite{wpssc14}  by using a criterion known as `edgeability'\cite{hmodl} . Any physical theory in strict $d$ spatial dimension must allow a physical edge to the vacuum while preserving symmetries. States that can only be realized at the surface of a $d+1$-dimensional SPT phase are clearly not edgeable. For the putative electron topological insulator with $\theta = 0$ and a fermionic monopole the edge theory was shown to be inconsistent within the electronic Hilbert space.

\subsection{Physics and construction of topological paramagnets}
\label{tp}
Having shown that the only root states not captured by band theory are the topological paramagnets, we now describe their physical properties.    The $Z_2^2$ classification means there are two root states.  Actually the group cohomology classification\cite{chencoho2011} gives a $Z_2$ classification, {\em i.e} it misses one of the two root states. This new root state was first proposed on physical grounds in Ref. \cite{avts12}.  Subsequently this was shown to exist using a coupled layer construction\cite{hmodl}, and using Walker-Wang models\cite{burnellbc}.

It is convenient to characterize these states by their representative surface states. If the symmetry is preserved then the surface realizes a quantum spin liquid state.  This quantum spin liquid must realize time reversal symmetry in a manner  forbidden in strictly $2d$ magnets.  Below we will describe two such surface quantum spin liquids for each of the two root states - one gapped and the other gapless. 

\subsubsection{Gapped surface quantum spin liquids}
The simplest such surface quantum spin liquid states have $\mathbb{Z}_2$ topological order, with two particles $e$ and $m$ having a mutual $\pi$-statistics. The first root SPT state supports a surface theory in which both $e$ and $m$ particles are Kramers bosons (denoted as $eTmT$), while the second SPT state has a surface in which both $e$ and $m$ are non-Kramers fermions ($e_fm_f$). The third state, being a composite of the previous two, has $e$ and $m$ both being Kramers fermions ($e_fTm_fT$). 

In all these quantum spin liquids the topological order itself naively seems consistent with time reversal symmetry but nevertheless they are not allowed in 
strictly $2d$ systems (see Supplementary Information). However these states can occur at the surface of $3d$ SPT states while preserving time reversal. This may be seen explicitly using the 
coupled layer construction\cite{hmodl}. For the $eTmT$ state we stack together 
layers of $Z_2$ quantum spin liquids such that in each layer $i$ the $e_i$ particle is Kramers doublet while $m_i$ is a Kramers singlet. This is allowed in strictly 2d systems, and indeed is the standard example\cite{ReSaSpN,wen91,z2long,bfg} of a $2d$ $Z_2$ quantum spin liquid. We then condense the combinations $e_{i+1}m_i e_{i-1}$. This confines all the bulk topological orders but leaves behind the $eTmT$ topological order at the surface. Similarly for the $e_fm_f$ state we stack together $2d$ $Z_2$ quantum spin liquids with trivial action of time reversal on the $e_i, m_i, \epsilon_i$. We then condense $\epsilon_{i+1} e_i \epsilon_{i-1}$.  This again confines all non-trivial particles in the bulk but leaves behind the $e_fm_f$ topological order on the surface. 

A different construction of the $e_fm_f$ state  was provided by Ref. \cite{burnellbc} through a powerful approach based on models introduced by Walker and Wang\cite{walkerwang} for $3d$ topological phases.   The key idea is to start with the $e_fm_f$ surface topological order and first construct a corresponding bulk ground state wave function in terms of superpositions of string configurations.    To that end three distinct species of strings are introduced that correspond to the three distinct particle types of the surface topological order. The $e_fm_f$ state when realized in strictly $2 +1$-D is described\cite{kitaev06,hmodl,burnellbc} by a time reversal breaking Topological Quantum Field Theory (TQFT) even though  the braiding and fusion rules themselves are time reversal symmetric.  In the three dimensional Walker-Wang wave function the amplitude for a string configuration $C$ is taken to be the same as the corresponding Wilson loop $W[C]$ in this $2+1$-D TQFT: 
\begin{equation}
\Psi_{3d}[C] = \langle W[C] \rangle_{2+1 ~TQFT}
\end{equation}

Formally this procedure can be used to write down $3d$ wavefunctions corresponding to any $2+1-D$ TQFT.  The different string types can fuse according to the fusion rules of this TQFT. In cases - like the $e_fm_f$ $Z_2$ topological order - where all particle types braid non-trivially around one another,  it is possible to show\cite{burnellbc} that the bulk $3d$ wave function is itself not topologically ordered. However when a $2d$ surface is introduced the boundary develops the intrinsic topological order of the $2+1-D$ TQFT. 
Note however that as all the braiding and fusion rules of the $2+1-D$ TQFT are time reversal symmetric the $3D$ ground state wave function $\Psi_{3D}$ is itself real (even with a boundary) and hence time reversal symmetric. 
Thus the Walker-Wang procedure leads us to the topological paramagnet with $e_fm_f$ surface topological order. Finally it is possible to construct a model for which $\Psi_{3D}$ is the exact ground state wave function\cite{burnellbc}. 

\subsubsection{Gapless surface quantum spin liquids}
\label{glessqsl}
We now describe a symmetry preserving gapless surface quantum spin liquid state for these root  topological paramagnets.  The equivalence to the gapped surface quantum spin liquid is shown in the Supplementary Section building on the results of Ref. \cite{TScSTO,wpssc14,wangts14}. For both root states the surface is again described by a deconfined $Z_2$ gauge theory. This theory has an electrically neutral fermionic Kramers doublet excitation $\chi$ (the spinon) that carries the $Z_2$ gauge charge. These fermions have a dispersion with  $8$ gapless Majorana cones, identical actually to that of {\em electrons} at the surface of a certain time reversal symmetric topological superconductor. At the free fermion level the gaplessness is thus protected by time reversal symmetry.  There is in addition a gapped vision excitation $v$ that carries the $Z_2$ gauge flux so that $\chi$ and $v$ have mutual $\pi$ statistics.   The two root states $eTmT$ and $e_fm_f$ differ in the statistics of $v$. In the former it is a boson while in the latter it is a fermion.

\subsection{Experimental fingerprints}
We now describe experimental fingerprints of the two new root states - the topological paramagnets. 
 The fingerprints of states obtained by taking combinations of these with each other or the TBI are readily inferred.

 If the topological paramagnets are in the gapless quantum spin liquid surface state, the gapless excitations can be probed through surface thermal transport experiments. 
  Alternately the surface may spontaneously break symmetry.  Detection of surface magnetism in a bulk paramagnet may then provide a hint of spin-SPT order. 
 More revealing fingerprints are obtained by breaking symmetry at the surface to obtain states with no intrinsic topological order.   The results are summarized in Table.\ref{root}.

\begin{figure}
 \includegraphics[scale=0.5]{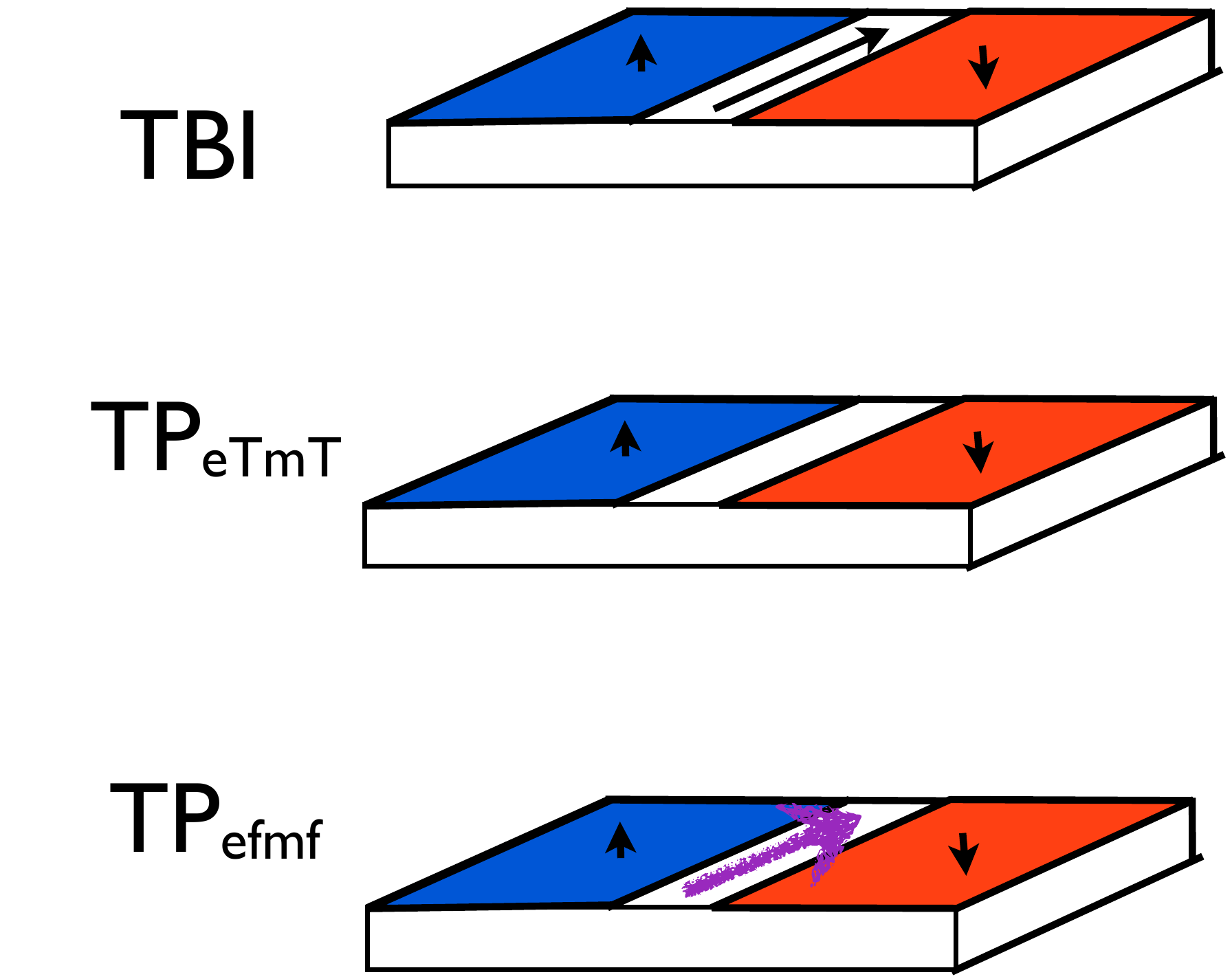}
\caption{Ferromagnetic domain wall at the surface of  the three root topological insulators. The TBI phase has a chiral charged edge mode in the domain wall. One Topological Paramagnet (labeled by surface topological $eTmT$) has a gapped domain wall while the other (labeled TP$_{efmf}$) has neutral chiral modes with $\nu_Q = 8$.}
\label{TIdomain}
\end{figure}

It is most useful to first consider depositing a ferromagnet at the surface to break time reversal. For all three root states the surface can then be gapped without any intrinsic topological order.  Let us now consider a domain wall where the magnetization changes sign (see Fig. \ref{TIdomain}). For the TBI this domain wall hosts a chiral electrically charged edge mode. For  the $e_fm_f$ topological paramagnet such a domain wall hosts gapless chiral modes but these are electrically neutral.  Furthermore these neutral modes transport heat as though there were eight species of neutral one-way fermions (more precisely the `chiral central charge' = 8).  These neutral modes may alternately be interpreted in terms of a thermal Hall conductivity of each domain of the form  $\kappa_{xy} = \nu_Q \frac{\pi^2}{3}\frac{k_B^2}{h} T$ with $\nu_Q = \pm 4$.  For the $e_Tm_T$  topological paramagnet on the other hand there are no gapless modes in such a domain wall, {\em i.e} for each domain $\kappa_{xy} = 0$.  

Thus a combined measurement of electrical and thermal Hall transport when ${\cal T}$ is broken at the surface can provide a very useful practical (albeit partial) characterization of these distinct topological insulators. 

Next we consider depositing an $s$-wave superconductor instead of a ferromagnet at the surface so that charge $U(1)$ symmetry is broken while preserving time reversal.     For the TBI this leads to a gapped surface but with exotic vortices that host Majorana zero modes\cite{fkmajorana}. For both the root states of the topological paramagnets it is convenient to start with the gapless surface spin liquid surface described above. The gapless Majorana fermion of the quantum spin liquid state can then hybridize with the Bogoliubov quasiparticle of the superconductor. This gets rid of the exotic excitations of the $Z_2$ quantum spin liquid. But now mixing between the gapless Majorana fermions and the Bogoliubov quasiparticles makes them visible to probes like Angle Resolved Photoemission (ARPES) or tunneling. Thus the induced superconductor will be gapless with $8$ Majorana cones that can be detected by ARPES.

 Taken together with the $T$-breaking surface transport we have a unique fingerprint for each of the 8 TIs. 
 
 \section{Correlated surface states for the $3d$ electronic TBI}
 \label{etisto}
 We now return to the $3d$ electronic TBI when it is realized in a strongly correlated system and ask about the fate of its surface states. Following the non-perturbative understanding of the surface of the boson TI, it is natural to ask if the surface of the electronic TBI can be gapped by strong interactions while preserving symmetry by paying the price of introducing intrinsic topological order.  
  
Two different such symmetry preserving surface topological orders were described in Refs. \cite{wpsprb13,maxetal,cfav13,bnq13}.  
Interestingly the topological order 
 at the surface of the electron TBI is necessarily {\em non-abelian}. 
By starting with the known superconducting surface\cite{fkmajorana} of the TBI and proliferating vortices it is possible to derive a surface topological order (dubbed T-Moore-Read) which is a time reversal symmetric version of the non-abelian Moore-Read fractional quantum Hall state\cite{wpsprb13,maxetal}.  A different non-abelian topological order (dubbed T-Pfaffian) was obtained through Walker-Wang methods\cite{cfav13} (and guessed in Ref. \cite{bnq13}) and shown to correspond to the surface of some electronic SPT state with   $\theta = \pi$ EM response.  Actually the methods of Ref. \cite{cfav13} yield two distinct realizations of time reversal symmetry in this surface topological order corresponding to two distinct bulk  SPT states. One of these describes the TBI and  the other its combination with the $eTmT$ topological paramagnet (the precise matching between these two is still not clear). 
 
Previously symmetric surface topological orders were also proposed for the time reversal invariant $3d$ topological superconductor\cite{TScSTO}. In some cases this is non-abelian (but of course different from the ones describing the TBI). In cases where it is abelian this proposed surface topological order can be derived simply by using a generalization of the vortex condensation arguments\cite{wangts14,fcmav14}. 

Two comments are appropriate: first as expected in all these cases the STO has anomalous implementation of the global symmetry: strictly $2d$ states with the same topological order cannot have the same symmetry implementation. Second, it is not granted that every $3d$ SPT phase admits such a symmetry preserving STO. Recent work\cite{wangts14} described 
an example (topological superconductor with spin $SU(2)$ and time reversal symmetries) where such a surface termination does not exist. In this case the surface is gapless so long as the symmetries are preserved even in the presence of strong interactions. This phenomenon was dubbed ``symmetry-enforced gaplessness".

\section{3d electronic SPTs with other symmetries}
\label{tsc3d}
The various kinds of arguments described above have enabled extensive progress\cite{wangts14} in describing and classifying $3d$ electronic SPT phases with various symmetries. Free fermion systems with various symmetries fall into one of 10 different classes (the 10-fold way) of topological phases\cite{tenfold}. With interactions we need to think about the specific symmetry group. For symmetry groups representative of each member of the 10-fold way, the classification and properties of SPT phases have been described\cite{wangts14}. These results are described in Table.  \ref{classtable}. We now highlight some of these results. 

 \begin{table*}[tttt]
\begin{tabular}{|>{\centering\arraybackslash}m{1.5in}|>{\centering\arraybackslash}m{1.2in}|>{\centering\arraybackslash}m{1in}|>{\centering\arraybackslash}m{1in}|}
\hline
{\bf Symmetry class} &  {\bf Reduction of free fermion states} & {\bf Distinct boson SPT} & {\bf Complete classification} \\ \hline
 $U(1)\rtimes\Z_2^T$ with $\T^2=-1$  (Spin-orbit coupled insulators)& $\Z_2\to\Z_2$ & $\Z_2^2$ & $\Z_2^3$ \\ \hline
\hline
$U(1)\times\Z_2^T$  (Triplet SC, conserved $S^z$) & $\Z\to\Z_8$ & $\Z_2$ & $\Z_8\times\Z_2$ \\ \hline
$(U(1)\rtimes\Z_2^{T})\times SU(2)$   (Spin-rotation invariant insulators)& 0 & $\Z_2^4$ & $\Z_2^4$ \\ \hline
\hline
$\Z_2^T$ with $\T^2=-1$  (Topological SC) & $\Z\to\Z_{16}$ & $0$ & $Z_{16}$ (?) \\ \hline
$SU(2)\times\Z_2^T$   (Spin-rotation invariant topological SC) & $\Z\to\Z_4$ & $\Z_2$ & $Z_4 \times Z_2$ (?) \\ \hline

\end{tabular}
\caption{A selection of results on classifications of electronic SPT states in three dimensions. More results and detail can be found in Ref. \cite{wangts14}. The second column gives free fermion states that remain nontrivial after introducing interactions. The third column gives SPT states that are absent in the free fermion picture, but are equivalent to those obtained from bosonic composites of the electron.  In the last 2 cases it is not currently clear if the classification is complete. 
}
\label{classtable}
\end{table*}%

For spin rotation invariant insulators with time reversal symmetry, within band theory there is no topological insulator. With interactions\cite{wangts14} the only SPT phases are bosonic SPTs formed by first forming Mott insulators and letting the resulting spins form a spin-SPT.  These are classified by $Z_2^4$ and have properties closely resembling time reversal symmetric spin systems with conserved $S^z$ discussed in Section \ref{bti3d}

Similarly for time reversal symmetric insulators of spinless fermions, the only SPT states are bosonic ones formed by SPTs of  neutral composites from the fermions\cite{wpssc14}. 

It is also interesting to consider the fate of topological superconductors in the presence of interactions.  The quadratic Bogoliubov-deGennes Hamiltonian for time reversal invariant superconductors in three dimensions allows for $Z$ classification corresponding to $n$ species of gapless Majorana fermions at the surface.  Using their Walker-Wang based studies of  surface topological order , Ref. \cite{TScSTO} argued that this reduces to a $Z_{16}$ classification in the presence of interactions. In other words when there are 16 species of gapless Majorana fermions at the surface the theory can be gapped out by interactions to yield a trivial surface (even though this does not happen in free fermion theory).  Refs.  \cite{wangts14,fcmav14} provided an elementary derivation of this result. 

Another interesting case is  to consider time reversal invariant triplet paired superconductors where $S^z$ is still conserved. The corresponding symmetry group is $U(1) \times Z_2^T$. In free fermion theory this has a $Z$ classification which reduces to 
a $Z_8 \times Z_2$ classification with interactions\cite{wangts14}. Here the $Z_8$ subgroup describes topological superconductors that admit a free fermion description while the $Z_2$ describes SPT states formed by bosonic composites of the electrons (or more precisely bosonic SPTs that are not equivalent to one of the free fermion topological SC states already contained in the $Z_8$ classification).

\section{Insights from SPTs for other problems}

 Though SPT phases are short range entangled (and hence fairly simple) their study has  given surprisingly powerful insights into the more complex `long range entangled' phases of matter. Here we briefly discuss the nature of these insights. 

{\bf Duality between SPT and topological order:} Perhaps the simplest long range entangled phases are those with intrinsic topological order  in $2d$ (like the fractional quantum Hall states or gapped $2d$ quantum spin liquids). 
Interestingly $2d$ SPT phases are related to these long ranged entangled phases through a duality transformation\cite{levingu}. A number of difficult theoretical questions (such as phase transitions) about such topologically ordered phases may thus be formulated as questions about SPT phases.

{ \bf Symmetry and intrinsic topological order:} The interplay of physical global symmetry and topological order is an old topic.  
For instance the possibility of fractional quantum numbers for quasiparticles (e.g., in the fractional quantum Hall effect  or for spinons in a quantum spin liquid)  is a statement about how  global  symmetries are implemented on the anyonic excitations of the associated topological order.  Recently topologically ordered phases in the presence of global symmetries have been christened ``Symmetry Enriched Topological" (SET) phases. 

What constraints are there about how symmetry may be implemented in topologically ordered states? Here the understanding of SPT phases provides powerful insights. Given some particular topological order with some implementation of symmetries, we may be able to generate others by putting one of the emergent quasiparticles in an SPT phase.  This enables us to establish a connection between seemingly different such topologically ordered phases. 

A different kind of constraint comes from the understanding of surface topological order in $3d$ SPT phases. The surface topological order has `anomalous' implementation of the global symmetry, {\i.e} strictly $2d$ systems with the same topological order cannot implement symmetry in the same way. Thus studying surface topological order generates `no-go' theorems on what kinds of symmetry implementations are legal in $2d$ topologically ordered states\cite{avts12,hmodl,statwitt,burnellbc,ctr14,xbvf14,kaptho14}. 

 The theory of SPT phases is thus a crucial ingredient in ongoing efforts to understand SET phases\cite{SET}.

{\bf Gapless quantum phases:} A number of frontier questions in correlated quantum systems relate to the physics of many body ground states with gapless excitations.  Perhaps the most familiar example to most physicists are phase of matter with a continuous global symmetry that is spontaneously broken. The broken symmetry gives rise to gapless Goldstone modes.  A different kind of protected gapless excitation is also  well known: For instance Landau quasiparticles in a Fermi liquid, or the Bogoliubov quasiparticles in a gapless BCS superconductor.  The protection of these gapless excitations as due to the {\em long range entanglement} in the corresponding ground state together with {\em unbroken} 
global symmetry.  Modern examples are the excitations of gapless quantum spin liquids and of non-fermi liquid metals. 

In the context of such gapless phases  an immediate question is whether they have symmetry realization that is legitimate in any strictly-$d$ dimensional system. In other words is the symmetry realization anomalous? Again studying possible gapless phases at the boundary of a $d+1$-dimensional  SPT phase generates no-go theorem for their existence in strictly 
$d$-dimensions.  For instance certain conjectured gapless quantum vortex liquid states\cite{avl} were recently shown\cite{hmodl} - based on insights from SPT phases - to not occur in strictly $2d$ magnets with time reversal symmetry. 

A different application of SPT ideas is to the physics of $3d$ quantum spin liquids with an emergent deconfined $U(1)$ gauge field (see Refs. \cite{bosfrc3d,hfb04,3ddmr,lesikts05,mlxgwrmp05,kdybk,shannon} for models with these phases) with a gapless photon.  Such phases are currently actively being sought in experiments on quantum spin ice materials on pyrochlore lattices\cite{balentsqspice}. Microscopic models of the spin interactions in these materials are complicated and have no internal symmetries other than time reversal. What can we say on general grounds about such time reversal invariant $U(1)$ quantum spin liquids? Using insights from the understanding of SPT phases it is possible to show\cite{cwtsunpub} that there are precisely 8 such distinct phases  where the only gapless excitation is the emergent photon. The properties of these phases are readily determined. Indeed many of these $8$ phases can be understood as gauged versions of either a boson or fermion SPT phase with both $U(1)$ and time reversal symmetry. 
The others can then be obtained as SPT phases of the either the emergent electric charge or the magnetic monopole. 
 
\section{Summary and Outlook}
SPT phases are a minimal generalization of the topological band insulators to interacting systems. We reviewed recent progress in our understanding of the physics of these phases. Our main focus was on three dimensional systems, particularly electrons,  with realistic symmetries.  We showed how a variety of physical arguments and methods can be fruitfully used to understand the possible interacting topological insulators and their physical properties.  

For systems with the symmetries  reviewed in this paper we now have a good conceptual understanding of the possible SPT phases and their universal physical properties.  
This progress however is just the beginning of what are likely to be bigger challenges. 
The immediate practical challenge for theory is to understand  the microscopic conditions that facilitate these phases. 
Clearly this will be aided by studying the realization of the $d > 1$ SPT phases in realistic models of particular experimental situations. In this context a theoretical question that has not received much attention so far is whether there exist sign-problem free Hamiltonians for $d > 1$ SPT phases and how SPT order might be detected in a quantum Monte Carlo simulation.  The results of Ref. \cite{lesik13}  on the $2d$ boson integer quantum Hall state are an encouraging sign that other SPT phases may be accessible in Monte Carlo calculations.  Another important question, which has just begun to be studied\cite{zaletel}, is 
 the identification of SPT order from the ground state wave function of some system for physically relevant symmetries. This may help detect SPT phases through DMRG calculations on strips.

There have thus far been very few suggestions for experimental realizations of these phases in $d > 1$. . For the boson integer quantum Hall state in $d = 2$, the model of two-component bosons at filling factor $\nu = 1$ each with delta function repulsion\cite{tsml12,ueda13,wujain13,nrts13} is a good guide. The challenge is to reach the quantum hall regime  in ultra-cold atoms. Other suggestions have been made for obtaining this phase in frustrated $2d$ lattice spin models in a strong magnetic field\cite{lulee12}. 
In $d = 3$ the topological paramagnets may be realized in electronic Mott insulators with frustrated spin interactions. Very recent work\cite{cwants14} suggests that a particularly good place may be in frustrated spin-$1$ quantum magnets in three dimensions.  

Some hints on physical realizations of bosonic SPT phases may be provided by a non-linear sigma model description developed in Refs. \cite{avts12,xu13,brx13}.  An interesting recent 
description of some of the interacting topological insulators as TBI phases of clusters of three electrons\cite{chongunpub} has been developed, and may also be a guide toward physical realization. 

On the conceptual side, an important open  issue is the phase transitions between various SPT phases. A start is in Refs. \cite{tgav12,luleeqpt12}. These are related to phase transitions of current interest\cite{maissam} in quantum spin liquid theory after some (or all) of the global symmetry is gauged. 

Very recently the braiding statistics description reviewed here for  $2d$ Ising SPTs has been generalized\cite{ml3loop14} to $3d$ for bosons with discrete global symmetry of the form $Z_n^k$. Interestingly this involves considering statistical phases associated with braiding of three distinct loops\cite{ml3loop14} - a concept that may also be useful in other $3d$ topological phases\cite{ran3lp14,wen3lp14}.

We described  some of the impact studies of SPT phases are having on other frontier problems in theoretical condensed matter physics.  
In a different direction, an interesting application\cite{wenso10,jwwen,ybx14} of SPT ideas is to issues of which chiral gauge field theories  (like the standard model of particle physics) can be lattice regulated while preserving all internal symmetries in the same space-time dimension.  This is a long standing question in lattice gauge theory studies of the standard model which may be clarified by insights from SPT physics.

\section{Acknowledgements}
 It is a pleasure to thank my collaborators M. Levin, A. Nahum, A. Potter, C. Xu, and particularly A. Vishwanath and Chong Wang. I also benefited immensely from discussions with 
 X. Chen, M. Metlitski, Matthew Fisher, Liang Fu, C. Kane, Z. Gu, and Xiao-gang Wen.  I also thank M. Levin for a careful reading of the paper, and the Perimeter Institute for Theoretical Physics for hospitality where part of the paper was written. This work was supported by NSF DMR-1305741, and partially supported by a Simons Investigator award from the Simons Foundation.

\appendix
\section{Anomalous symmetry at the topological paramagnet surfaces}
Let us ask if the $eTmT$ or $e_fm_f$ states are edgeable.  In both these states the $\epsilon$ particle is a Kramers singlet.   It will be convenient to consider a putative edge theory where the $\epsilon$ is tuned to be gapless. A time reversal symmetric system cannot be chiral, and hence the number $n_R$ of right-moving channels of $\epsilon$ particles at the edge must equal the number $n_L$ of left-moving channels. Denoting these right/left moving (Majorana) fermions $\chi_{Ra}$ ($\chi_{La}$) the corresponding edge Hamiltonian may be written
\begin{equation}
\label{h1d}
H_{1d} = \int dx \sum_{a=1}^{n} \chi_{Ra} \left(-i\partial_x \right) \chi_{Ra} + \chi_{La} \left(i\partial_x \right) \chi_{La}
\end{equation}
Under time reversal we have $\chi_{Ra} \rightarrow \chi_{La}$ and vice versa.

It is useful to have a bulk model which has this edge structure. Start initially with two species of bulk fermions $\epsilon_+$ and $\epsilon_-$. We put $\epsilon_{\pm}$ in $n$ copies of a  $p_x \pm ip_y$ paired state. This system is time reversal invariant if $\epsilon_{+} \rightarrow \epsilon_{-}$ (and vice versa) under time reversal.  The well known edge structure of the $p_x \pm ip_y$ paired states ensure that the edge Hamiltonian has the form Eqn. \ref{h1d}. 
In the bulk we then add all possible (local) time reversal invariant terms. These will include inter-species mixing terms, in particular mixing between $\epsilon_+$ and $\epsilon_-$. 
Now in the bulk the $e$ (and $m$) particle  is a mutual semion with $\epsilon$. If we ignore interspecies mixing,  the $e$ particle has $n$ Majorana zero modes of $\epsilon_+$ and likewise $n$ Majorana modes for $\epsilon_-$.  However inter-species mixing will produce lift the degeneracy associated with these zero modes. The result is a unique Kramers singlet bosonic $e$ particle (and similarly for the $m$ particle).  Thus neither the $e_fm_f$ nor the $eTmT$ state is realized. 
 
Now let us understand this generally from  the edge theory. The $e$ and $m$ can be identified with 'twist' operators that shift the phase of $\epsilon$ by $\pi$.  The (abelian) statistics of the bulk $e$ particle is determined by the `spin'  of the corresponding operator at the edge ({\em i.e} the difference of the scaling dimension in the right and left moving theories).  As the right and left movers are related by time reversal it follows that the twist operator has $0$ spin - hence the bulk $e$ particle is a boson. Thus the time reversal symmetric $e_fm_f$ state is not edgeable, and cannot be realized in strict $2d$.  Indeed it is easy to see\cite{kitaev06} that in strict $2d$ $e$ will be a fermion if $n_R - n_L = 4$ mod $8$ (and hence requires broken time reversal).

To address the edgeability of the $eTmT$ state let us begin by considering $n_R = n_L = 1$. Then the edge theory is the same as the critical Ising model. A twist operator is the Ising order parameter field $\sigma$ which we take to be the edge avatar of the $e$ particle. Under time reversal as $R$ and $L$ interchange, the right and left moving components of $\sigma$ interchange but clearly $\sigma \rightarrow \sigma$. In particular $\sigma$ is a Kramers singlet. If we have $n$ species of edge Majorana fermions, the $e$ particle may be represented as $\sigma_1 \sigma_2.......\sigma_n$ where $\sigma_a$ is the Ising order parameter for species $a$. Clearly this too just goes to itself under time reversal and is not a Kramers doublet. Thus we conclude that the $eTmT$ topological ordered state  is not edgeable, and hence cannot be realized in any strictly $2d$ magnet.  

Note that the $eTmT$, $e_fm_f$  topological orders are related through a phase transition to the gapless $Z_2$ quantum spin liquids described in Section \ref{glessqsl}.  Thus we conclude that these gapless spin liquids can also not be realized in strictly $2d$ magnet. A corollary is that the $8$ gapless Majorana modes of this state cannot be trivially gapped while preserving time reversal. If such gapping were possible we could then condense the bosonic vision $v$ (present in the $eTmT$ case)  and get a trivial gapped state.

\section{Gapless quantum spin liquids at the surface of topological paramagnets}
We begin with an observation of 
 Ref. \cite{TScSTO} about the surface topological order of   {\em electronic} topological superconductors with ${\cal T}$ symmetry.  Using Walker-Wang techniques it was argued that when the surface has $8$ gapless Majorana cones it can be gapped to leave behind either $eTmT$ or $e_fm_f$ surface topological order. An elementary argument to understand this was described in Refs. \cite{wpssc14} which we now sketch.  We first group the $8$ gapless Majorana fermions into $4$ Dirac fermions with the Lagrangian
 \begin{equation}
\mathcal{L}_{free}=\sum_{i=1}^{4}\psi^{\dagger}_i(p_x\sigma^x+p_y\sigma^z)\psi_i,
\end{equation}
in which time-reversal acts as
\begin{equation}
\mathcal{T}\psi_i\mathcal{T}^{-1}=i\sigma_y\psi^{\dagger}_i.
\end{equation}
It is easy to see that there is no time reversal symmetric quadratic mass term that can gap these fermions. To address it non-perturbatively we first enlarge the symmetry to $U(1) \times Z_2^T$ with the $U(1)$ acting as 
\begin{equation}
U_{\theta}\psi U^{-1}_{\theta}=e^{i\theta}\psi
\end{equation}
We then add a `pairing' mass term  
\begin{equation}
\label{gap}
\mathcal{L}_{gap}=i\Delta\sum_{i=1}^{4}\psi_i\sigma_y\psi_i+h.c.
\end{equation}
which breaks both $U(1)$ and $\mathcal{T}$ but preserves the combination $\tilde{\mathcal{T}}=\mathcal{T}U_{\pi/2}$. We now attempt to restore the broken symmetry while preserving $\tilde{{\mathcal T}}$ by proliferating vortices of  the pair order parameter. However the vortices have zero modes which restricts the kinds of vortices that can condense.  With $8$ Majorana cones, the single vortex can be shown to be a Kramers doublet under $\tilde{\mathcal{T}}$, and hence cannot condense. 
 The ``minimal" construction is to proliferate double vortices. The resulting insulating state has $\mathbb{Z}_2$ topological order $\{1,e,m,\epsilon\}$ where we identify the $e$ particle with the unpaired vortex and the $\epsilon$ with the `neutralized'  version of the fermionic quasiparticle.  
 
Now the full $U(1)\times\mathcal{T}$ is restored, we can ask how are they implemented on $\{1,e,m,\epsilon\}$. Obviously these particles are `charge'-neutral and the question is then about the implementation of $\mathcal{T}$ alone. However, since the particles are `neutral'  the extra auxiliary $U(1)$ rotation in $\tilde{\mathcal{T}}$ is irrelevant and they transform identically under $\tilde{\mathcal{T}}$ and $\mathcal{T}$. Thus $\mathcal{T}^2=\tilde{\mathcal{T}}^2=-1$ on $e$ and $m$, and $\mathcal{T}^2=\tilde{\mathcal{T}}^2=1$ on $\epsilon$, which is exactly the topological order $eTmT$. The physical `charged' fermion $\psi$ is now trivially gapped and plays no role in the topological theory.  We can now introduce explicit pairing to break the auxiliary $U(1)$ symmetry. Since topological order stems from the `charge-neutral' sector, pair-condensation of $\psi$ does not alter the topological order, and the resulting state is just the $eTmT$ state with only $\mathcal{T}$ symmetry.

Applying this logic now to the proposed gapless $Z_2$ quantum spin liquids, we see that these states are connected through surface phase transitions to the gapped state 
$(1, e_T, m_T, \epsilon) \times (1, \chi, v, v\chi)$.  Now if the $v$ is a boson we cam simply condense it to leave behind the $e_Tm_T$ topological order. If on the other hand $v$ is a fermion we can condense the combination $v\epsilon$. It is easy to see that this leaves behind the $e_fm_f$ topological order. This establishes the promised equivalence between the gapped and gapless surface quantum spin liquids.

\appendix

\end{document}